\documentclass[floatfix,aps,prl,twocolumn,showpacs,preprintnumbers,amsmath,amssymb]{revtex4}
\usepackage{epsfig}
\usepackage{graphicx}
\usepackage{dcolumn}
\usepackage{bm}
\usepackage{ulem}
\usepackage{color}

\begin{document}
\title{Determination of the Superconductor-Insulator Phase Diagram for One-Dimensional Wires}
\author{A. T. Bollinger$^\ast$, R. C. Dinsmore III, A. Rogachev$^\dag$, and A. Bezryadin}
\affiliation{Department of Physics, University of Illinois at
Urbana-Champaign, Urbana, Illinois 61801-3080, USA}
\date{\today}

\begin {abstract}
We establish the superconductor-insulator phase diagram for
quasi-one-dimensional wires by measuring a large set of MoGe
nanowires. This diagram is consistent with the
Chakravarty-Schmid-Bulgadaev phase boundary, namely with the
critical resistance being equal to $R_Q = h/4e^2$. We find that
transport properties of insulating nanowires exhibit a weak Coulomb
blockade behavior.
\end {abstract}

\pacs{74.78.Na, 74.25.Dw, 74.40.+k}

\maketitle

In quasi-one-dimensional superconductors it remains to be fully
understood how the superconducting properties of a wire are
destroyed as its diameter is reduced. Years ago it was shown that
for wires with micron size diameters the mechanism that weakens
superconductivity at finite temperature, $T$, are thermally
activated phase slips (TAPS), which break the superconducting phase
coherence along the wire and result in a measurable resistance, $R$
\cite{LAMHetal, TinkhamBook}. As temperature is reduced the thermal
fluctuations that cause TAPS freeze out and the TAPS rate decreases
faster than exponentially until at $T = 0$ these phase slips are
absent from the wire altogether and it should be in a truly
superconducting state ($R = 0$). However, in the ultrathin wires
being fabricated today this simple picture is complicated by an
additional phase breaking process due to quantum fluctuations
\cite{Mooij, Giordano, GZQPS}. As $T \rightarrow 0$ these quantum
phase slips (QPS) remain active and the resistance of a wire remains
finite, even at $T = 0$. Since the free energy barrier to phase
slips is proportional to the wire's cross sectional area, the
thinner a wire is made the more readily phase slips should occur in
it and therefore the QPS resistance is higher \cite{Lau, Arutyunov,
Altomare}.

While it would seem that ultrathin superconducting wires loose the
beneficial property of dissipationless electrical transport, the
remarkable possibility exists to recover the truly superconducting
state if QPS are suppressed. Recent experiments on a group of six
wires with similar lengths $\sim$100 nm observed that as $T
\rightarrow 0$ those wires whose normal state resistance, $R_N$, was
less than some critical resistance, $R_c$, were superconducting,
while wires with $R_N \gtrsim R_c$ were resistive, with increasing
resistance as $T \rightarrow 0$ \cite{Bezryadin}. It was found that
$R_c \approx R_Q$, where $R_Q = h/4e^2$, which is suggestive of a
Chakravarty-Schmid-Bulgadaev (CSB) dissipative phase transition
\cite{CSB}, in which QPS can be inhibited due to the interaction
with a dissipative environment. This transition was originally
predicted for shunted Josephson junctions but recently theoretically
generalized for thin wires \cite{Buchler, RefaelJJ, RefaelRT,
RefaelNew}. Unfortunately, these early experiments could not provide
a proof of the universality of the condition $R_c = R_Q$.
Furthermore, it is not clear whether a real superconductor-insulator
transition (SIT) occurs in ultrathin superconducting wires or merely
a crossover from wires in which the QPS rate is too small to be of
consequence and so appear superconducting to wires in which the QPS
rate is so large they essentially drive the wire into the normal
state. Distinguishing between these two possibilities is of critical
importance not only to our understanding of the physics of
quasi-one-dimensional superconducting wires but also to their
applicability in miniaturized superconducting circuits
\cite{TinkhamAPL}.

In this Letter, we present results obtained on a large collection of
about 100 wires that provide definitive evidence for the SIT and
show that for short wires the phase boundary is the same as in the
CSB transition, suggesting the same physical mechanism. No
indication of a crossover caused by a gradual increase of the QPS
rate was found. The wires have been characterized by linear
transport measurements as well as high-bias differential resistance
measurements. The results allow us to sort most of the homogeneous
samples into two categories, ``superconducting'' and ``insulating'',
and to construct a phase diagram with a well defined boundary for
the SIT in thin wires. This diagram will provide an experimental
basis for the theory of thin superconducting wires, which is still
being developed \cite{GZQPS, Buchler, RefaelRT, RefaelNew, Sachdev,
Khlebnikov}.

The nanowires used in this study were fabricated by molecular
templating \cite{Bezryadin}, by depositing amorphous \cite{Graybeal}
Mo$_{0.79}$Ge$_{0.21}$ (sputtered MoGe thicknesses were in the range
4-16 nm) onto insulating fluorinated single-wall carbon nanotubes
that were suspended over a trench etched into Si/SiO$_2$/SiN
substrates. The wires were carefully examined by scanning electron
microscopy (SEM) and the choice of the wire for transport
measurements was influenced by such factors as apparent homogeneity,
straightness of the wire, desired dimensions, and good connection of
the wire to the films on either side of the trench (SEM measured
lengths and widths of the wires were in the ranges 29-490 nm and
8.8-40.0 nm, respectively). Transport measurements were performed in
$^4$He and $^3$He cryostats, and some samples were further tested in
a top-loading dilution refrigerator. Each system had $\pi$ and
microwave filtering placed at room- and low-temperature stages,
respectively.

\begin{figure}[b]
\begin{center}
\includegraphics[width = 2.8in]{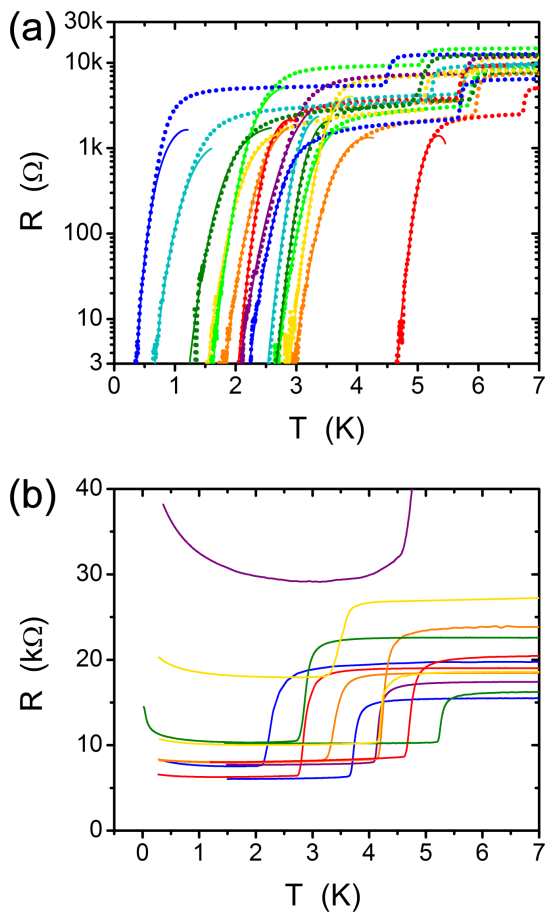}
\caption{\label{fig:RvsT} (a) $R$ vs. $T$ data ($\bullet$) for wires
in the superconducting phase with fits to the TAPS theory (lines).
(b) $R$ vs. $T$ for wires in the insulating phase.}
\end{center}
\end{figure}

The phase to which a nanowire belongs is easily discerned by the
transport properties of the wire.  In Fig.\ \nolinebreak
\ref{fig:RvsT} we show the $R(T)$ curves for some representative
samples. Note that upon cooling all samples initially show a
superconducting transition at the critical temperature of the thin
film electrodes, $T_{c,film}$, that are measured in series with the
wire. For $T < T_{c,film}$ the resistance of the electrodes is zero
and we probe only the wire. The normal state resistance of a wire,
$R_N$, is assumed to be the sample resistance just below
$T_{c,film}$.

Superconducting wires (Fig.\ \nolinebreak \ref{fig:RvsT}(a)) show
resistive transitions that are well described by the TAPS theory
(details of the fitting procedure can be found in Ref.\ \nolinebreak
\cite{RogachevPRL1}; alternate methods are given in Refs.\
\nolinebreak \cite{RefaelRT, KhlebnikovRT}). We emphasize that for
wires in the superconducting phase no resistance ``tails'' or other
non-TAPS behavior that could possibly be attributed to QPS are
observed, even for those wires near the SIT. However, one should be
careful in interpreting this result. According to the microscopic
theory of Zaikin et al. \cite{GZQPS}, the contribution to resistance
from QPS, $R_{QPS} \propto \text{exp}(-aR_QL/R_N\xi(0))$ where
$\xi(0)$ is the zero-temperature coherence length and $L$ is the
length of the wire, contains a numerical factor, $a$, that is of
order one and depends on the actual dependence of the order
parameter phase on time and space coordinates during the QPS
process. We find that setting this coefficient to $a \geq 4.44$ is
enough to suppress the expected contribution from QPS below the
level of noise in the experiments, and so free QPS, if they do
indeed occur in the superconducting wires, can not be resolved.
While this value for $a$ is different from the value of $a = 0.8$
found in Ref. \cite{Markovic}, it is still of order one as required
by the theory. But this is not the end of the story since one also
needs to test the same expression with the same ``a'' on the
insulating wires.

The $R(T)$ curves of wires in the insulating phase (Fig.\
\nolinebreak \ref{fig:RvsT}(b)), with resistance that increases upon
cooling, are clearly different from those in the superconducting
phase. One way to understand the insulating regime would be to
consider it qualitatively the same as the superconducting regime,
except that the QPS rate is higher due to the smaller diameter of
the wires. In such a crossover model, the QPS occur in all wires,
all wires retain some non-zero resistance at zero temperature, and
the QPS parameter $a$ is the same for all wires. However, this model
does not explain the abruptness of the transition observed: The
wires are either in agreement with the TAPS model or show an
insulating behavior.  We did not observe any intermediate regime
with a mixture of TAPS and QPS contributions even though a large set
of wires was studied. Reentrant behavior, i.e. resistance increasing
with cooling and then suddenly dropping, was never observed as well.
Also, such a crossover model quantitatively contradicts the value $a
= 4.44$ discussed above. The predicted resistance from QPS at $T =
0$ with $a = 4.44$ for some of the typical insulating wires is less
than 1\% of $R_N$ (and frequently many orders of magnitude less than
this), which is never observed (the observed value is always
slightly greater than $R_N$ (Fig.\ \nolinebreak \ref{fig:RvsT}(b))).
We found that there is no single value of the coefficient $a$ that
can explain the whole set of data. Thus, we suggest that the correct
model for quasi-one-dimensional superconducting nanowires should
involve an SIT, with the insulating phase characterized by complete
elimination of the superconducting condensate, due to proliferating
QPS.

\begin{figure}[b]
\begin{center}
\includegraphics[width = 2.9in]{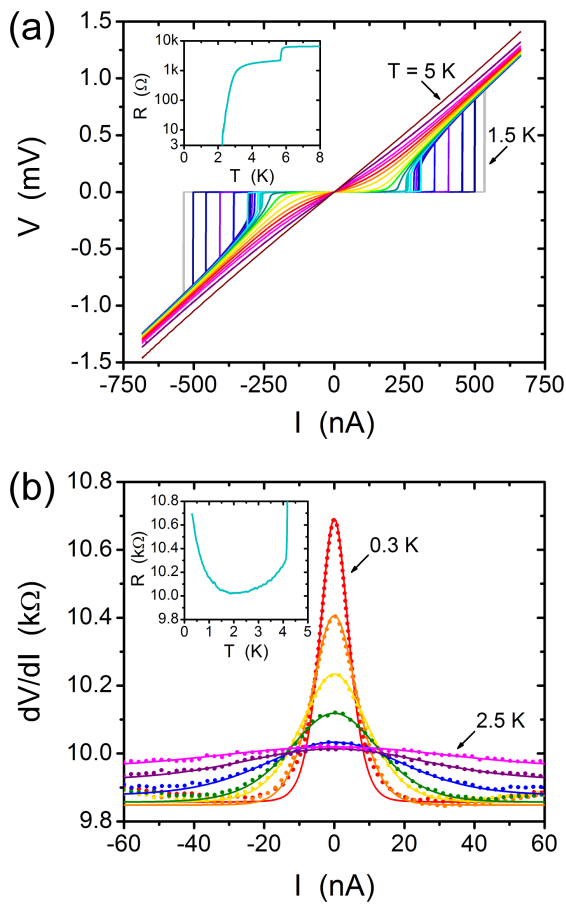}
\caption{\label{fig:VvsI} (a) $V$ vs. $I$ for a representative wire
in the superconducting phase ($L =$ 68 nm, $T_c =$ 4.02 K) at $T =$
1.5, 1.6, 1.7, 1.8, 1.9, 2.0, 2.1, 2.2, 2.3, 2.4, 2.6, 2.8, 3.0,
3.25, 3.5, 4.0, and 5.0 K. (Inset) $R$ vs. $T$ for the same sample.
(b) $dV/dI$ vs. $I$ data ($\bullet$) for a representative wire in
the insulating phase at $T =$ 0.3, 0.5, 0.75, 1.0, 1.5, 2.0, and 2.5
K with fits to Coulomb blockade theory (lines). (Inset) $R$ vs. $T$
for the same sample.}
\end{center}
\end{figure}

Another dichotomy in transport properties is also found in the
voltage vs. current, $V(I)$, characteristics of the wires. In Fig.\
\nolinebreak \ref{fig:VvsI}(a), $V(I)$ curves for a representative
superconducting sample at different temperatures show the evolution
of the $V(I)$ behavior of wires in this phase from linear for $T >
T_c$, to nonlinear for $T_c > T \gtrsim T_c/2$, to hysteretic for $T
\lesssim T_c/2$ with well defined switching and retrapping currents.
Insulating wires, on the other hand, display $V(I)$ characteristics
that are nearly linear at all temperatures but with a zero-bias
anomaly that is more pronounced in the differential resistance,
$dV(I)/dI$, measurements. For a representative insulating sample we
show $dV(I)/dI$ data at different temperatures in Fig.\ \nolinebreak
\ref{fig:VvsI}(b). Since a small zero-bias maximum can be observed
even for those temperatures at which $R(T)$ is at its minimum, the
$dV(I)/dI$ peak cannot be a result of Joule heating. Instead, as
shown by the fits in Fig.\ \nolinebreak \ref{fig:VvsI}(b), the
zero-bias anomaly can be described by the theory of dynamical weak
Coulomb blockade with the entire wire acting as a coherent
scatterer, the same as for the $R(T)$ curves for these wires
\cite{BollingerEPL, NGZCS}.

\begin{figure}[b]
\begin{center}
\includegraphics[width = 2.8in]{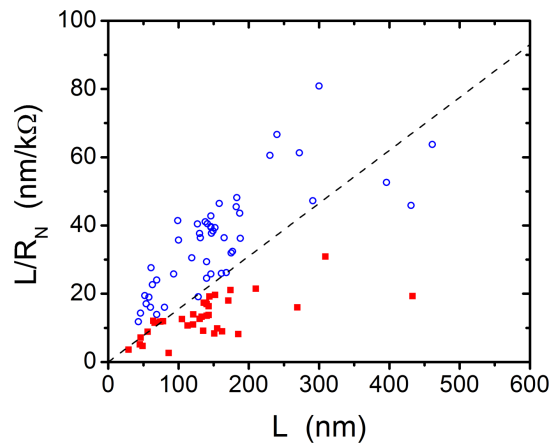}
\caption{\label{fig:LoRnvsL} Phase diagram of all superconducting
(\textcolor{blue}{$\circ$}) and insulating (\textcolor{red}{{\tiny
$\blacksquare$}}) wires in $L/R_N$-$L$ space. Dashed line is $R_N =
R_Q$.}
\end{center}
\end{figure}

We now turn our attention to the main result of this Letter - the
phase diagram of the SIT in quasi-one-dimensional nanowires. Based
upon the transport measurements the wires fall into one of two
distinct phases: superconducting or insulating \cite{EndNote1}. In
Fig.\ \nolinebreak \ref{fig:LoRnvsL} we plot the phase to which a
wire belongs on the coordinate plane ($L$,$L/R_N$). If the SIT in
ultrathin wires is caused by local physics then there should exist a
critical cross sectional area, $A_c$, that separates insulating and
superconducting wires. For the range of MoGe thicknesses sputtered,
the resistivity does not change with thickness \cite{Graybeal} and
so the separatrix in this scenario should appear on Fig.\
\nolinebreak \ref{fig:LoRnvsL} as a horizontal line with $L/R_N =
constant$. This is certainly not the phase boundary we observe.
Rather, the CSB phase boundary, i.e. $R_N = R_Q$ (dashed line in
Fig.\ \nolinebreak \ref{fig:LoRnvsL}) provides a much better
division of the data. It is observed that three longer wires
($\sim$450 nm) behave as superconductors even though they have high
normal resistance, i.e. $R_N > R_Q$. These deviations suggest that
the SIT is only applicable to shorter wires, as predicted in Ref.\
\nolinebreak \onlinecite{RefaelNew}.

\begin{figure}[b]
\begin{center}
\includegraphics[width = 2.8in]{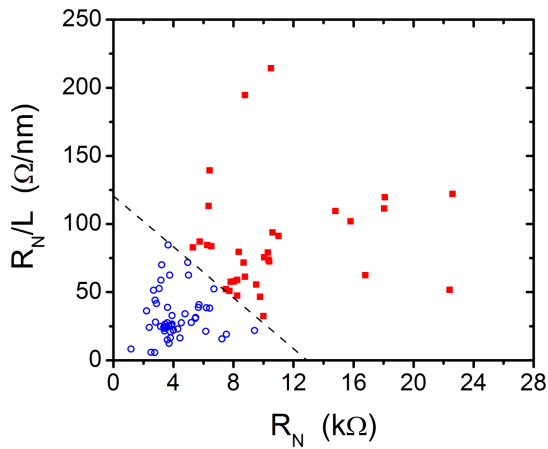}
\caption{\label{fig:RnoLvsRn} Phase diagram of all superconducting
(\textcolor{blue}{$\circ$}) and insulating (\textcolor{red}{{\tiny
$\blacksquare$}}) wires in $R_N/L$-$R_N$ space with the separatrix
(dashed line) given in the text.}
\end{center}
\end{figure}

Some short wires with $R_N$ somewhat lower than $R_Q$ appear
insulating. These deviations can be explained by assuming that our
knowledge about the effective $R_N$ is not precisely correct. The
effective $R_N$ might in fact go above the measured $R_N$ as the
temperature is reduced. A proximity effect can also be responsible
for the imprecise knowledge of the effective $R_N$ of the wire.
However, these deviations may point to other phenomena occurring in
these wires. In Fig.\ \nolinebreak \ref{fig:RnoLvsRn} we plot the
state of wires on the ($R_N$,$R_N/L$) coordinate plane. In this
representation it is clear that a single line can be drawn that
separates precisely the superconducting and insulating phase. The
phase boundary is given by $R_N/L = (2R_Q - R_N)/107$ nm. Samples
below the line are superconducting while those above it are
insulating.  A more useful form of this boundary is obtained by
using $R_N = \rho_NL/A$, where $A$ is the cross sectional area of
the wire, and the typical value for the normal metal resistivity of
MoGe, $\rho_N =$ 180 $\mu\Omega$-cm \cite{Graybeal}. Thus the
separatrix simply is $A = A_c$ where $A_c = 14.9$ nm$^2/(1 -
R_N/2R_Q)$. The superconducting (insulating) phase occurs for
samples with $A > A_c$ ($A < A_c$). This means that for wires that
have $R_N \ll R_Q$ the superconductivity is lost if the wire
diameter, $d$, is less than the critical diameter  $d_c \sim 4.4$
\nolinebreak nm. This critical diameter is quantitatively consistent
with our recent conjecture that superconductivity in ultrathin wires
is affected, in accordance with the Abrikosov and Gor'kov mechanism,
by magnetic moments that spontaneously form on the wire surface
\cite{RogachevPRL2}. Assuming that magnetic pair breaking is
responsible for the destruction of superconductivity in MoGe wires
with $R_N \ll R_Q$ one can estimate the critical diameter from the
empirical law relating wire diameter to the exchange scattering
time, $\tau_{B}$, found in Ref.\ \nolinebreak \cite{RogachevPRL2}.
The critical exchange scattering time, below which $T_c$ is zero, is
given by $\tau_{B,c} = \hbar S (S + 1) / \alpha_c = 3 \hbar
e^{\gamma} / 2 \pi k_B T_{c0}$ where $S = 1/2$, $\alpha_c$ is the
critical depairing factor, $\gamma = 0.577$ is Euler's constant, and
$T_{c0}$ is the critical temperature of the wire in the absence of
pair breaking effects. In Ref. \cite{RogachevPRL2}, a fit to
$\tau_{B}$ vs. $d$ data for MoGe nanowires showed that $d \sim 3$
\nolinebreak nm/ps $\times \ \tau_{B}$ and $T_{c0}$ was found to be
in the range $4.4 - 5.6$ \nolinebreak K. This corresponds to
$\tau_{B,c} \sim 1.2 - 1.5$ \nolinebreak ps and $d_c \sim 3.5 - 4.4$
\nolinebreak nm, in agreement with the value of $d_c$ obtained from
Fig.\ \nolinebreak \ref{fig:RnoLvsRn}. Finally, we point out that
the empirical phase boundary in Fig.\ \nolinebreak
\ref{fig:RnoLvsRn} suggests that wires with $d \gg d_c$ will become
insulating if $R_N > 2R_Q$, i.e. when localization effects become
strong.

In conclusion, we have studied a large set of nanowire samples with
lengths and normal state resistances in the ranges of 29-490 nm and
1.17-32.46 k$\Omega$, respectively. The phase diagram of the SIT is
in good, albeit not exact, agreement with the one expected for the
CSB transition, in accordance with the theory of Ref.\ \nolinebreak
\onlinecite{RefaelNew}. The few deviations can be accounted for by
the destruction of superconductivity due to local magnetic moments
in wires that would otherwise belong to the superconducting part of
the diagram. This comparison with the CSB phase boundary assumes
that $R_N$ can be considered as an effective shunting resistance as
validated by Ref.\ \nolinebreak \onlinecite{RefaelNew}.

\begin{acknowledgments}
We thank M.W. Brenner, E. Demler, G. Refael, M. Sahu, and T.-C. Wei
for assistance and discussions. The work was supported by the U.S.
Department of Energy, Division of Materials Sciences under Award No.
DEFG02-91ER45439, through the Frederick Seitz Materials Research
Laboratory at the University of Illinois at Urbana-Champaign and the
NSF CAREER Grant DMR 01-34770.
\end{acknowledgments}



\begin{thebibliography}{99}
\bibitem{EndNote1}
As was demonstrated in Ref.\ \nolinebreak \cite{BollingerPRB},
inhomogeneous samples usually can not be characterised decisively as
superconducting or insulating. Thus we did not include such samples
on the SIT diagram. Also, among the samples that did not show clear
signs of inhomogeneity, two samples showed a behavior intermediate
between superconducting and insulating.  We explain this by probable
hidden inhomogeneity as an atypical time lag occured between
fabrication and measurement for these two samples. We also exclude
them from the diagram.
\end{thebibliography}
\end{document}